\documentclass[prd,twocolumn,showpacs,amsmath,amssymb]{revtex4}
\usepackage{amssymb}
\usepackage{mathrsfs}
\usepackage{txfonts}

\usepackage{graphicx}
\usepackage{dcolumn}
\usepackage{bm}

\begin{document}

\title{Consistency of the tachyon warm inflationary universe models}

\author{Xiao-Min Zhang}
\email{zhangxm@mail.bnu.edu.cn}
\affiliation{Department of Physics, Beijing Normal University, Beijing 100875, China}
\author{Jian-Yang Zhu}
\thanks{Corresponding author}
\email{zhujy@bnu.edu.cn}
\affiliation{Department of Physics, Beijing Normal University, Beijing 100875, China}
\date{\today}

\begin{abstract}
This study concerns the consistency of the tachyon warm inflationary models. A linear stability analysis is performed to find the slow-roll conditions, characterized by the potential slow-roll (PSR) parameters, for the existence of a tachyon warm inflationary attractor in the system. The PSR parameters in the tachyon warm inflationary models are redefined. Two cases, an exponential potential and an inverse power-law potential, are studied, when the dissipative coefficient $\Gamma=\Gamma_0$ and $\Gamma=\Gamma(\phi)$, respectively. A crucial condition is obtained for a tachyon warm inflationary model characterized by the Hubble slow-roll (HSR) parameter $\epsilon_{_H}$, and the condition is extendable to some other inflationary models as well. A proper number of e-folds is obtained in both cases of the tachyon warm inflation, in contrast to existing works. It is also found that a constant dissipative coefficient $(\Gamma=\Gamma_0)$ is usually not a suitable assumption for a warm inflationary model.
\end{abstract}
\maketitle

\section{\label{sec:level1}Introduction}
With the inflationary phase added to the standard Big Bang model, many long-standing problems (horizon, flatness and monopoles) can be solved naturally \cite{Kazanas,Sato,Guth1981,Linde1982,Albrecht1982}. It is generally agreed that the inflation produced seeds that give rise to the large scale structure and to the observed little anisotropy of cosmological microwave background (CMB) \cite{WMAP}. During the standard inflation, which is sometimes called cold inflation, the Universe undergoes a steep supercooling phase, for it is assumed that the scalar field which is responsible for the inflation is isolated and the interaction between the inflaton and other fields are neglected. After the supercooling phase, the Universe needs a reheating epoch to get hot again and get filled with radiation required by the Big-Bang model. The density perturbation, which is the seed of the structure formation, is mainly due to the quantum fluctuation. Another type of the inflation named warm inflation was first proposed in \cite{BereraFang}. In that scenario, there are interactions between the inflaton and the other fields, and $\mathcal{L}_{int}$ in the Lagrangian density of the scalar field describes
the interaction of inflaton with all other fields \cite{BereraFang}. The friction term including $\Gamma\dot\phi$ in the equation of motion of the inflaton phenomenologically describes the decay of the inflaton field into the thermal bath via the interaction Lagrangian $\mathcal{L}_{int}$ \cite{BereraFang,Berera1996}. The Universe is hot, and the radiation production occurs constantly while the Universe accelerates. There is no need for a reheating epoch, and the connection with the radiation dominated Big-Bang phase is smooth. But the new inflationary scenario was criticized by Yokoyama and Linde \cite{YokoyamaLinde} for one cannot get large enough number of e-folds and the viscosity term is negligible from quantum field theory. In these early works such as \cite{YokoyamaLinde}, dissipation effects were being looked at in a high temperature regime, and it proved too difficult to keep finite
temperature effective potential corrections to be small. At almost the same time, Berera, Gleiser and Ramos proposed a more complicated model \cite{BereraGleiser} to make warm inflation possible, and much problem was eliminated by invoking supersymmetry in the picture \cite{BereraRamos2001,BereraRamos2003}. A successful two-stage interaction configuration was proposed in \cite{BereraRamos2001} to make warm inflation work. In that configuration, the inflaton was coupled to heavy catalyst fields with masses larger than the temperature of the Universe, and these fields in turn were coupled to light fields. The first analysis of dissipative coefficients for the two-stage mechanism was done by Moss and Xiong in \cite{MossXiong}. Some recent papers \cite{Mar2011,Mar2013} on calculating the dissipation coefficients also examined the warm inflation from quantum field theory. A significant feature of the warm inflation is that the density perturbations originate from the thermal fluctuations \cite{BereraFang,Lisa2004,Berera2000}. The ¡°$\eta$-problem¡± \cite{etaproblem} and the overlarge amplitude of the inflaton suffered in the cold inflation \cite{Berera2005,BereraIanRamos} can be eliminated in the warm inflation. Moreover, as the slow-roll conditions in warm inflation were suggested, a broader potential can be used to realize warm inflation. As in the cold inflation, the scalar field which drives the inflation is also an inflaton field and the potential of the inflaton is also the dominating energy during the inflation. Matter in the Universe can be generated by the decay of the inflaton field or the radiation field \cite{Taylor1997}.

The tachyon field might be responsible for the cosmological inflation at a very early Universe \cite{reheatingproblem,TachyonWI,SetareKamali,XiongLi,XiaoZhu}, and can be a candidate for the dark matter \cite{Sami2002} and the dark energy \cite{Copeland2005}. The tachyon field is associated with unstable D-branes in string theory \cite{ASen2002,ASen2006}, and has a Lagrangian density with a non-canonical kinetic term ($\mathcal{L}_{tach}=-V(\phi)\sqrt{1-\partial^{\mu}\phi\partial_{\mu}\phi}$), which is a generalization of the Lagrangian of a relativistic particle \cite{ASen2006}. As illustrated in \cite{SetareKamali}, the tachyonic inflation is a type of k-inflation. It is interesting to combine warm inflationary scenario with tachyon field, and these attempts have been considered in some works, such as \cite{TachyonWI,SetareKamali,XiaoZhu}. However, the consistency problems, such as whether the slow-roll assumption is reasonable and thermal correction is small enough, have not been checked and are specifically considered in this paper. Generally speaking, the tachyon potentials have two properties: it reaches maximum at $\phi\rightarrow 0$, and minimum at $\phi\rightarrow \infty$. We will analyze two types of potentials that satisfy these two conditions: an exponential potential ($V(\phi)=V_0e^{-\alpha\phi}$) and an inverse power law potential ($V(\phi)=C\phi^{-m}$).

The inflation is often associated with a slow-roll approximation that neglects the highest order terms in the dynamical equations of the system. We shall consider whether the slow-roll equations can describe the inflation of the Universe exactly. A stability analysis is often used to obtain the conditions for the system to remain close to the slow-roll solution for many Hubble times. The stability analysis for a canonical scalar field in the warm inflation can be found in Refs. \cite{Ian2008,Campo2010}. We will perform a linear stability analysis for tachyon field in the warm inflation to get the slow-roll conditions. In our stability analysis, we redefine the PSR parameters to make the calculation concise. We should also point out that the number of e-folds calculated in some papers \cite{TachyonWI,XiaoZhu} is less than one. We shall provide correction to some of what we believe are mistakes in their calculations, and give the correct number of e-folds for the tachyon warm inflation. Based on our analysis of the two example potentials, we obtain a crucial condition for a inflationary model, characterized by the HSR parameter $\epsilon_{_H}$, and we extend it to some other inflationary models. We also find that a constant dissipative coefficient ($\Gamma=\Gamma_0$) is not a good assumption for the two example potentials.

The paper is organized as follows. The next section contains a brief introduction to the tachyon warm inflationary Universe. Sec. \ref{sec:level3} analyzes the slow-roll conditions in the tachyon warm inflation. The two example potentials, the exponential form and the inverse power-law form, are studied in Sec. \ref{sec:level4} and Sec. \ref{sec:level5}, respectively. Finally, Sec. \ref{sec:level6}
contains discussions and conclusions.

\section{\label{sec:level2}Tachyon warm inflationary Universe}
 A rolling tachyon matter in a spatially flat Friedmann-Robertson-Walker (FRW) Universe is described by an effective fluid with the energy-momentum tensor $T_{\nu}^{\mu}=diag(-\rho_{\phi},p_{\phi},p_{\phi},p_{\phi})$ \cite{Gibbons2002}, where the energy density $\rho_{\phi}$ and the pressure $p_{\phi}$ for a tachyon field are defined by
 \begin{equation}\label{rhophi}
    \rho_{\phi}=\frac V{\sqrt{1-\dot\phi^2}} ,
 \end{equation}
 and
\begin{equation}\label{pphi}
p_{\phi}=-V\sqrt{1-\dot\phi^2} ,
\end{equation}
respectively. In the equations above, $\phi$ denotes the tachyon field and $V$ the effective potential associated with the tachyon field. From the potential properties we know that $V$ should satisfied $V,_{\phi}<0$, where the subscript $\phi$ denotes a derivative with respect to $\phi$.

The Friedmann equation and the equation of motion for the inflaton in the FRW cosmological model are given by
\begin{equation}\label{Friedmann}
H^2=\frac{1}{3M_p^2}\rho,
\end{equation}
and
\begin{equation}\label{EOM}
\frac{\ddot\phi}{1-\dot\phi^2}+3H\dot\phi+\frac{V,_{\phi}}{V}=-\frac{\Gamma}{V}\sqrt{1-\dot\phi^2}\dot \phi ,
\end{equation}
respectively, where $H=\dot a/a$ is the Hubble factor, $a$ is the FRW scale factor, $M_p^2=(8\pi G)^{-1}$, and $\rho$ is the total energy density of the multi-component system of tachyon and radiation. Equation (\ref{EOM}) can be derived from the conservation equation
\begin{equation}\label{conservation}
\dot\rho_{\phi}+3H(\rho_{\phi}+p_{\phi})=-\Gamma\dot\phi^2,
\end{equation}
where $\Gamma$ is the dissipative coefficient ($\Gamma>0$ by the Second Law of Thermodynamics) and is responsible for the decay of the inflaton into a thermal bath during the inflation. Unlike cold inflation, here the Universe has a finite temperature, because the radiation production occurs simultaneously with the inflationary
expansion. The components of the universe during inflation are the tachyon field and the radiation, and the total energy density $\rho$ and the pressure $p$ are given by \cite{Ian2008,Lisa2004}
\begin{equation}\label{totalrho}
\rho=\frac{V(\phi,T)}{\sqrt{1-\dot\phi^2}}+Ts\simeq\rho_{\phi}+\rho_r,
\end{equation}
and
\begin{equation}\label{totalp}
p=-V(\phi,T)\sqrt{1-\dot\phi^2}\simeq p_{\phi}+\frac13\rho_r,
\end{equation}
 where $\rho_r$ is the radiation energy density. We should note here that the total energy density can be written in a separable form of inflaton energy and radiation energy, owing to the slow-roll condition we shall obtain in Sec. \ref{sec:level3} for the slow-roll parameter $b$. In the inflationary regime, where the slow-roll conditions hold, the finite-temperature effective potential $V(\phi,T)$ has the form $V(\phi,T)\simeq V(\phi)+V(T)$, where $V(\phi)$ is the tachyon potential in Eq. (\ref{rhophi}), and $V(T)$ contributes to the radiation energy. From the total energy-momentum conservation equation, $\dot\rho+3H(\rho+p)=0$, and Eq. (\ref{conservation}), we can get
\begin{equation}\label{rhor}
\dot\rho_r+4H\rho_r=\Gamma \dot\phi^2,
\end{equation}
Or, equivalently,
\begin{equation}\label{entropy}
T\dot s+3HTs=\Gamma \dot\phi^2,
\end{equation}
when a thermal correction to the effective potential is negligible (which means $\rho_r=\frac34 Ts$ and the condition of $s\propto T^3$ is satisfied). The slow-roll approximation in tachyon warm inflation means $\dot\phi^2\ll1$ and $\ddot\phi\ll(3H+\Gamma/V)\dot\phi$, which is very different from the conditions of warm inflation for a
 canonical field, $\dot\phi^2\ll V$ and $\ddot\phi\ll(3H+\Gamma)\dot\phi$. It is reasonable to consider the potential energy of tachyon field to be the dominating energy during the inflation ($\rho \sim V$), and the production of the radiation to be quasi-stable. Under these assumptions and the slow-roll conditions, the Friedmann equation and the motion equation of the inflaton are reduced to
\begin{equation}\label{SRFriedmann}
H^2=\frac V{3M_p^2},
\end{equation}
and
\begin{equation}\label{SREOM}
3H(1+r)\dot\phi+\frac{V,_{\phi}}{V}=0,
\end{equation}
where $r=\Gamma/3HV$ is the parameter that characterizes the strength of the dissipative effect, with $r\gg1$ for the strong dissipative regime and $r\ll1$ for the weak dissipative regime. Equations (\ref{rhor}) and (\ref{entropy}) can be reduced to
\begin{equation}\label{SRentropy}
3HTs=\Gamma \dot\phi^2,
\end{equation}
and
\begin{equation}\label{SRrhor}
4H\rho_r=\Gamma\dot\phi^2.
\end{equation}
In thermodynamics, we have $\rho_r=\sigma T^4$, where $\sigma$ is the Stefan-Boltzmann constant.

Two HSR parameters are defined as
\begin{equation}\label{HSRepsilon}
\epsilon_{_H}=-\frac{\dot H}{H^2}=\frac{M_p^2}{2}\frac{1}{1+r}\frac{V,_{\phi}^2}{V^3},
\end{equation}
and
\begin{equation}\label{HSReta}
\eta_{_H}=-\frac{\ddot H}{H\dot H}\simeq \frac{M_p^2}{(1+r)V}\left[\frac{V,_{\phi\phi}}{V}-\frac12\left(\frac{V,_{\phi}}{V}\right)^2\right].
\end{equation}
The warm inflation can take place when the condition $\epsilon_{_H}<1$ (implies $\ddot a >0$) holds. The HSR parameter $\epsilon_{_H}$ is an important parameter for estimating whether the inflation will last forever. When $\epsilon_{_H}=1$, the inflationary phase ends. The number of the e-folds during the tachyon warm inflation is
\begin{equation}\label{efold}
N=-\frac{1}{M_p^2}\int^{\phi_e}_{\phi_*}\frac{V^2}{V,_{\phi}}(1+r)d\phi,
\end{equation}
where $\phi_e$ denotes the inflaton when inflation ends, and $\phi_*$ the Hubble horizon crossing.

\section{\label{sec:level3}Stability analysis}
Inflationary solutions to the exact equations (\ref{Friedmann}), (\ref{EOM}) and (\ref{entropy}) are difficult to obtain, and a slow-roll approximation is often introduced. The slow-roll approximation involves $\dot\phi^2\ll1$ and neglecting the highest order terms in the exact equations. But question remains as to under what conditions can the slow-roll equations (\ref{SRFriedmann}) - (\ref{SRentropy}) describe the system well. We shall perform a linear-stability analysis to obtain the conditions for the system to remain close to the slow-roll solutions for many Hubble times, i.e., the slow-roll solution should be an attractor for the dynamical system. For convenience, we define a new variable $u=\dot\phi$, so that $\ddot\phi=\dot u$. Now Eqs. (\ref{EOM}) and (\ref{entropy}) can be rewritten as
\begin{equation}\label{dotu}
\frac{\dot u}{1-u^2}+3Hu+\frac{V,_{\phi}}{V}=-\frac{\Gamma}{V}\sqrt{1-u^2}u,
\end{equation}
and
\begin{equation}\label{dotentropy}
T\dot s+3HTs=\Gamma u^2.
\end{equation}
Through Eqs. (\ref{Friedmann}) and (\ref{totalrho}), we obtain
\begin{equation}\label{dotH}
2H\dot H=\frac{1}{3M_p^2}\left(V,_{\phi}u+Vu\dot u+T\dot s\right),
\end{equation}
where we have used the condition of $u^2\ll1$. Then we obtain the rate of change of the Hubble parameter as
\begin{equation}\label{HbyH2}
\frac{1}{H}\frac{d\ln H}{dt}=\frac{\dot H}{H^2}=-\frac32\frac{Vu^2+Ts}{V+Ts}.
\end{equation}
During the slow-roll inflationary period, the Hubble parameter is nearly constant, which means $|\dot H/H^2|\ll1$. From Eq. (\ref{HbyH2}), we get $u^2\ll1$ and $Ts\ll V$, which are consistent with the assumptions of $\dot\phi^2\ll1$ and the potential domination during inflation.

We should note that, in our stability analysis, we do not use the such approximations as $u^2\ll1$. in order to check the self-consistency of the slow-roll assumptions which are often used in previous papers about tachyon inflation. For the stability analysis, we define three small perturbation variables $\delta\phi$, $\delta u$ and $\delta s$, along with background variables $\phi_0$, $u_0$, and $s_0$, which denote the slow-roll solutions that satisfy
\begin{equation}\label{u0}
3H_0(1+r_0)u_0=-\frac{V,_{\phi}}{V_0},
\end{equation}
\begin{equation}\label{T0s0}
3H_0T_0s_0=\Gamma_0u_0^2,
\end{equation}
and
\begin{equation}\label{H02}
H_0^2=\frac{V_0}{3M_p^2}.
\end{equation}
The exact solutions can be expanded as $\phi=\phi_0+\delta\phi$, $u=u_0+\delta u$, and $s=s_0+\delta s$. We assume that the perturbation terms are much smaller than the background ones, i.e., $\delta\phi\ll\phi_0$, $\delta u\ll u_0$, and $\delta s\ll s_0$.

We also redefine the PSR parameters, which should be straightforward since they only contain the potential, the dissipative coefficient, and the derivative of the potential and the dissipative coefficient with respect to the inflaton field. The new PSR parameters will make the stability analysis concise and easy. The new definitions of the PSR parameters in the tachyon warm inflation are
\begin{equation}\label{newPSR}
\tilde{\epsilon}=\frac{M_p^2}{2}\frac{V,_{\phi}^2}{V^3},~~~\tilde{\eta}=M_p^2\frac{V,_{\phi\phi}}{V^2},~~~
\tilde{\beta}=M_p^2\frac{V,_{\phi}\Gamma,_{\phi}}{V^2\Gamma}.
\end{equation}
The other two parameters describing the temperature dependence are the same as in the non-tachyon warm inflation
\begin{equation}\label{bc}
b=\frac{TV,_{\phi T}}{V,_{\phi}},~~~~c=\frac{T\Gamma,_T}{\Gamma}.
\end{equation}
By using $s\simeq -V,_T$, we get
\begin{equation}\label{deltas}
\delta s=-V,_{TT}\delta T-V,_{\phi T}\delta\phi.
\end{equation}
We have the formula of $Ts,_{_T}=3s$ when the thermal correction to the potential is small, and using the formula we get
\begin{equation}\label{deltaT}
\delta T=\frac1{3s_0}\left(T_0\delta s+V,_{\phi}b\delta\phi\right).
\end{equation}
Taking the variation of Eq. (\ref{Friedmann}), we get
\begin{eqnarray}
2H_0\delta H &=&\frac 1{3M_p^2}\left[ \left( \frac{V,_\phi }{\sqrt{1-u_0^2}}-%
\frac{V,_\phi b}{3\sqrt{1-u_0^2}}+\frac{V,_\phi b}3\right) \delta \phi \right.
\nonumber \\
&&\left. +V u_0(1-u_0^2)^{-3/2}\delta u+\left( \frac 43T-\frac T{3\sqrt{1-u_0^2}}%
\right) \delta s\right] . \nonumber \\
\label{deltaH}
\end{eqnarray}
Similarly, we get the variation of $V$, $V_{\phi}$, and $\Gamma$
\begin{equation}\label{deltaV}
\delta V=\left(V,_{\phi}-\frac13 V,_{\phi}b\right)\delta\phi-\frac13 T_0\delta s,
\end{equation}
\begin{equation}\label{deltaVphi}
\delta V_{\phi}=\left(\frac{V_0\eta}{M_p^2}+\frac{V,_{\phi}^2b^2}{3s_0T_0}\right)\delta\phi+\frac{V,_{\phi}b}{3s_0}\delta s,
\end{equation}
\begin{equation}\label{deltagamma}
\delta\Gamma=\Gamma_0\left(\frac{V_0\beta}{M_p^2V,_{\phi}}+\frac{V,_{\phi}bc}{3s_0T_0}\right)\delta\phi+
\frac{c\Gamma_0}{3s_0}\delta s.
\end{equation}
These formulas will be used thereinafter.

For convenience, we express the equations of small variations of the perturbations in a matrix form
\begin{equation}  \left(
\begin{array}{c} \delta \dot \phi \\ \delta \dot u \\  \delta \dot s
\end{array} \right)  =E \left(\begin{array}{c} \delta \phi \\
 \delta u \\  \delta s \end{array} \right) -F,     \end{equation}
where $E$ is a $3\times3$ matrix
\begin{equation}
E=\left(\begin{array}{ccc} 0&1&0\\ A&\lambda_1 &B\\ C&D&\lambda_2
\end{array} \right),
\end{equation}
and the matrix element can be written as
\begin{eqnarray}
A &=&3H_0^2\left\{ \left( 1+r_0\right) bcM^{3/2}-\frac{\left(
1+r_0\right) ^2}{r_0}b^2M-\tilde{\eta}M\right.   \nonumber \\
&&+\frac{r_0}{1+r_0}\tilde{\beta}M^{3/2}+\frac 1{1+r_0}\tilde{\epsilon}%
M^{1/2}\left[ 1-\frac b3-2r_0M\right.   \nonumber \\
&&\left. \left. +\frac 23br_0M+\left( 2+2r_0-\frac b3-\frac 23br_0\right)
M^{1/2}\right] \right\}
\end{eqnarray}
\begin{eqnarray}
B &=&\frac{H_0T_0}{u_0V_0}\left[ -cM^{2/3}+\frac{1+r_0}{r_0}bM-%
\frac{2\tilde{\epsilon}}{3(1+r_0)}M\right.   \nonumber \\
&&\left. -2u^2M+\frac{u^2}2M^{1/2}-r_0u^2M^{3/2}\right] ,
\end{eqnarray}
\begin{eqnarray}
C &=&\frac{3H_0^2u_0V_0}{T_0}\left[ \frac{r_0}{2(1+r_0)}\tilde{\epsilon}-%
\frac{r_0}{1+r_0}\tilde{\beta}\right.   \nonumber \\
&&\left. +(1+r_0)(1-c)b\right] ,
\end{eqnarray}
\begin{equation}
D=\frac{H_0u_0V_0}{T_0}\left[ 6r_0-\frac{r_0}{(1+r_o)^2}\tilde{\epsilon}%
\right] ,
\end{equation}
\begin{eqnarray}
\lambda _1 &=&-3H_0M-3H_0r_0M^{3/2}-H_0\frac{\tilde{\epsilon}}{%
\left( 1+r_0\right) ^2}M^{-1/2}  \nonumber \\
&&+3H_0r_0u\left( M^{1/2}+2uM^{1/2}-2u\right) ,
\end{eqnarray}
\begin{equation}
\lambda _2=-H_0(4-c)-H_0\frac{r_0\tilde{\epsilon}}{\left( 1+r_0\right) ^2}.
\end{equation}
$M=1-u_0^2$ in the above equations.
We have used Eqs. (\ref{dotu}), (\ref{dotentropy}), and (\ref{deltaH}) - (\ref{deltagamma}) for obtaining the formulas.

The column matrix $F$ is a small `` forcing term'', which can be expressed as
\begin{equation}
F=\left(\begin{array}{c} 0\\ \dot u_0 \\ \dot s_0 \end{array}
\right) ,
\end{equation}
The slow-roll solution can be an attractor for tachyon warm inflation only when the the ``forcing term'' is small enough and the matrix $E$ have negative eigenvalues. We shall study this ``forcing term'' $F$ first. Taking the derivative of Eqs. (\ref{u0}) and (\ref{T0s0}) with respect to time, we get
\begin{equation}
\dot u_0=\frac{\tilde{B}C-\tilde{A}\lambda_2}{\tilde{\lambda_1}\lambda_2-\tilde{B}D}u_0,
\end{equation}
and
\begin{equation}
\dot s_0=\frac{\tilde{A}D-C\tilde{\lambda_1}}{\tilde{\lambda_1}\lambda_2-\tilde{B}D}u_0.
\end{equation}
The parameters $\tilde{A},\tilde{B},\tilde{\lambda_1}$ are given by
\begin{eqnarray}
\tilde{A} &=&3H_0^2\left[ (1+r_0)bc-\frac{(1+r_0)^2}{r_0}b^2-\tilde{\eta}\right.
\nonumber \\
&&\left. +\frac{r_0}{1+r_0}\tilde{\beta}+\frac{3-2/3b}{1+r_0}\tilde{\epsilon}%
\right] ,
\end{eqnarray}
\begin{equation}
\tilde{B}=\frac{H_0T_0}{u_0V_0}\left[-c+\frac{1+r_0}{r_0}b-\frac{2}{3(1+r_0)}\tilde{\epsilon}\right],
\end{equation}
\begin{equation}
\tilde{\lambda_1}=-3H_0(1+r_0)-H_0\frac{\tilde{\epsilon}}{(1+r_0)^2},
\end{equation}
Using the expressions of the associated matrix element and the expressions for $\tilde{A},\tilde{B},\tilde{\lambda_1}$ and neglecting non-linear terms of the PSR parameters, we can finally get
\begin{equation}
\frac{\dot{u}_0}{Hu_0}=\frac 1\Delta \left[ 3(1+r_0)bc+\frac{4r_0}{1+r_0}%
\tilde{\beta}+(c-4)\tilde{\eta}+\frac{r_0c-6}{2(1+r_0)}\tilde{\epsilon}%
\right] ,  \label{dotu0H0u0}
\end{equation}
and
\begin{eqnarray}
\frac{\dot{s}_0}{Hs_0} &=&\frac 3\Delta \left[ 2(1+r_0)bc-2\tilde{\eta}+%
\frac{r_0-1}{r_0+1}\tilde{\beta}\right.   \nonumber \\
&&\left. +\frac{3+r_0}{2(1+r_0)}\tilde{\epsilon}+\frac{(1+r_0)^2(1-c)}{r_0}%
b\right] .  \label{dots0H0s0}
\end{eqnarray}
Using the equations above and Eq. (\ref{deltaT}), we get
\begin{eqnarray}
\frac{\dot{T}_0}{HT_0} &=&\frac 1\Delta \left[ 2(1+r_0)bc-2\tilde{\eta}+%
\frac{r_0-1}{r_0+1}\tilde{\beta}+\frac{3+r_0}{2(1+r_0)}\tilde{\epsilon}%
\right.   \nonumber \\
&&\left. -\frac{(1+r_0)^2}{r_0}(2cr_0+3r_0+3)b\right] ,  \label{dotT0H0T0}
\end{eqnarray}
where
\begin{equation}\label{Delta}
\Delta=(4+c)r_0-2(1+r_0)b.
\end{equation}
Using the HSR and the PSR parameters, we get the relationship for $\dot H_0/H_0^2$
\begin{equation}\label{dotH0H02}
\epsilon_{_H}=-\frac{\dot H_0}{H_0^2}=\frac{\tilde{\epsilon}}{1+r}\ll1.
\end{equation}
The validity of the slow-roll approximation requires the rate of change of Hubble parameter and the ``forcing term'' to both be small. A small ``forcing term'' is equivalent to $|\dot u_0/Hu_0|\ll1$ and $|\dot s_0/Hs_0|\ll1$. These conditions can be satisfied if the PSR parameters
\begin{equation}
\tilde{\epsilon}\ll 1+r,\quad |\tilde{\eta}|\ll 1+r,\quad |\tilde{\beta}|\ll
1+r,\quad |b|\ll \frac r{1+r}.  \label{SR}
\end{equation}
which are obtained by using Eqs. (\ref{dotu0H0u0}), (\ref{dots0H0s0}), and (\ref{dotH0H02}). Since $b$ is much smaller than other slow-roll parameters,
\begin{equation}\label{Delta1}
\Delta\simeq(4+c)r_0+4-c.
\end{equation}
From the analysis, we see that the slow-roll condition $u^2\simeq\frac{2\tilde{\epsilon}}{3(1+r)^2}\ll1$ is satisfied by $\tilde{\epsilon} \ll 1+r$, which means that the slow-roll assumption is satisfied in the slow-roll regime, hence the self-consistency. In the slow-roll regime, $M\rightarrow 1$ and then the parameters $A,B,\lambda_1$ can be reduced to $\widetilde{A},\widetilde{B}, \widetilde{\lambda_1}$. An interesting finding here is that the tachyon behaves like a canonical field in the slow-roll regime. One can see this through a field redefinition of $\psi=\int \sqrt{V} d\phi$ when $u^2\ll1$ and $b\ll1$. But tachyon field evolves differently from canonical field at late time. The parameter $\tilde{\epsilon}=1+R$ is equivalent to $\epsilon_{_H}=1$, which implies the end of the inflationary phase.
Next we shall study the eigenvalues of the matrix $E$. Using the PSR conditions we have obtained, we find the matrix element $A$ and $C$ are much smaller than the others. The characteristic equation for $E$ is
\begin{eqnarray}
det(\lambda I-E) &\simeq &\left|
\begin{array}{ccc}
\lambda  & -1 & 0 \\
0 & \lambda-\lambda_1  & -B \\
0 & -D & \lambda-\lambda_2
\end{array}
\right|   \nonumber \\
&=&\lambda (\lambda-\lambda _1 )(\lambda-\lambda _2 )-BD\lambda   \nonumber
\\
&=&0.  \label{charEq}
\end{eqnarray}
There exists a small eigenvalue
\begin{equation}
\lambda \simeq \frac{-BC-A\lambda _2}{\lambda _1\lambda _2-BD-A}\ll\lambda_1,\lambda_2.
\label{charlambda}
\end{equation}
$\lambda_1,\lambda_2 $ satisfy the equation
\begin{equation}
\lambda ^2-(\lambda _1+\lambda _2)\lambda +\lambda _1\lambda _2-BD=0.
\label{twoeigen}
\end{equation}
Both eigenvalues are negative only when $\lambda _1+\lambda _2<0$ and $\lambda
_1\lambda _2-BD>0$. Using the expressions of the matrix elements, in the limit of $M\rightarrow1$ we get
\begin{equation}
|c|<4.  \label{c}
\end{equation}
We have obtained all the conditions for the slow-roll parameters in potential form. Using the newly defined PSR parameters, we finally get the stability conditions in the same form as that of the canonical scalar field constructed in terms of the traditional PSR parameters (i.e., $\epsilon =  M_p^2V,_{\phi}^2/2V^2$, $\eta =M_p^2V,_{\phi\phi}/V$, $\beta =M_p^2 V,_{\phi} \Gamma, _{\phi}/V\Gamma $) \cite{Ian2008,Lisa2004}. The conditions obtained above for the slow-roll parameters $b$ and $c$ are the same as in the non-tachyon scalar field case \cite{Ian2008}. We should note that the condition for $b$ implies that the slope of the thermal correction to the effective potential has to be small. The condition on parameter $c$ only means the temperature dependence of the dissipative coefficient should be within the range of $\Gamma\propto (T^{-4}, T^4)$.

Next we shall study two cases of potential, both satisfing the two properties of tachyon potential \cite{SetareKamali,TachyonWI} in the strong dissipative regime. Since the condition on $b$ guarantees that the thermal correction to the effective potential is small, we can write the total energy in a separable form
\begin{equation}\label{separablerho}
    \rho=\frac{V(\phi)}{\sqrt{1-\dot\phi^2}}+\rho_r.
\end{equation}

\section{\label{sec:level4}exponential potential in the strong dissipative regime}

Let us consider a tachyon field with the potential of
\begin{equation}\label{eVphi}
V(\phi)=V_0e^{-\alpha\phi},
\end{equation}
where $V_0$ and $\alpha$ are positive free parameters. $V_0$ is in unit of $M_p^4$, and $\alpha$ is related to the tachyon mass \cite{Fairbarin2002} with unit of $M_p$. We should note here that tachyon is a kind of non-canonical field \cite{Franche2010} and the Lagrangian density of tachyon are often written in the form of $\mathcal{L}=-V\sqrt{1-\dot\phi^2}$ in homogeneous FRW background, which is not a uniform normalization form that can reduce to canonical case (i.e. $\mathcal{L}=X-V$, where $X=\frac12\dot\phi^2$) in small $X$ limit. Thus the tachyon field has dimension $[\phi]=\mathbf{M}^{-1}$ in natural unit where $c=1$, and $\mathbf{M}$ denotes the dimension of mass. If we make a uniform normalization of the tachyon field $\psi=\int \sqrt{V} d\phi$, then the rescaled field $\psi$ has the normal dimension $[\psi]=\mathbf{M}$ as the usual scalar field. Our recent work about non-canonical warm inflation \cite{zhangzhu2} found that, since the tachyon field is not in a uniform normalization form, the dissipation coefficient $\Gamma$ in the tachyon warm inflationary model is quite different from that (denote as $\tilde{\Gamma}$) in canonical scalar field or uniform normalized non-canonical scalar field. The dimension of dissipation coefficient $\tilde{\Gamma}$ is $[\tilde{\Gamma}]=\mathbf{M}$ as usual, thus the dissipation coefficient $\Gamma\sim\tilde{\Gamma}V$ in the tachyon warm inflationary case has dimension of $[\Gamma]=\mathbf{M}^5$. Therefore the rate parameters $r$ in the canonical case ($r=\Gamma/3H$) and the tachyon case ($r=\Gamma/3HV$) can both be dimensionless number. In canonical scalar field or uniform normalized non-canonical scalar field, $\tilde{\Gamma}$ is often set to a constant, $\tilde{\Gamma}=\tilde{\Gamma}_0$, for simplicity\cite{BereraFang,Herrera2010}, thus in our case $\Gamma\propto V$. Some works, however, assume that $\Gamma=\Gamma_0=constant$ in the tachyon warm inflationary case, thus $\tilde{\Gamma}\propto V^{-1}$, which seems quite unreasonable. In this and the next sections, we shall prove that $\Gamma\propto V$ is a better choice than $\Gamma=\Gamma_0$. This kind of exponential potential has been widely used in cosmological inflation theories \cite{Lucchin,Halliwell,Ratra}, and can result in power-law inflation. Moving beyond conventional inflation using canonical scalar field, since exponential potential satisfies the two properties of tachyon field, it is natural to use it in tachyon inflation \cite{TachyonWI,XiongLi,XiaoZhu}. This exponential potential does not have a minimum, as opposed to many non-tachyon field potentials. The warm inflation does not need a reheating phase where the inflaton oscillates about the minimum, and is a good mechanism for tachyon to act as the inflaton.

We will restrict ourselves to the strong dissipative regime ($r\gg1$) in the following.

\subsection{\label{sec:level41}$\mathbf{\Gamma=\Gamma_0}$ case}

Here we assume the dissipative coefficient is a constant $\Gamma_0$.
The newly defined PSR parameters are
\begin{equation}\label{newPSRforGamma_0}
\tilde{\epsilon}=\frac{\alpha^2M_p^2}{2V},~~~\tilde{\eta}=\frac{\alpha^2M_p^2}{V}=2\tilde{\epsilon},~~~\tilde{\beta}=0.
\end{equation}
The two parameters describing the temperature vanish in this case. The Hubble parameter and the rate $r$ are given by, respectively,
\begin{equation}\label{Hphi}
H(\phi)=\frac1{\sqrt{3}M_p}\sqrt{V_0}e^{-\alpha\phi/2},
\end{equation}
and
\begin{equation}\label{rater}
r=\frac{\Gamma_0 M_p}{\sqrt{3}V_0^{3/2}}e^{3\alpha\phi/2}.
\end{equation}
The energy density of the radiation field is
\begin{equation}\label{energyrho}
\rho_r=\frac{\Gamma u^2}{4H}\simeq\frac{\tilde{\epsilon}}{2r}\rho_{\phi}.
\end{equation}
We should point out that, in \cite{TachyonWI}, the authors obtained the total number of the e-folds $N_{total}=[1-(V_e/V_i)^{1/2}]$. The total number of the e-folds $N_{total}$ is less than one due to what believe is a mistake in their calculation. In fact this is the number of e-folds before the inflation ($V_e$ is the potential at the beginning of inflation). The number of e-folds before the inflation is less than one and can be ignored. We shall correct this in a distinct way.

The HSR parameters $\epsilon_{_H}$ and $\eta_{_H}$ are given by, respectively,
\begin{equation}\label{epsilonH}
\epsilon_{_H}=\frac{\sqrt{3}M_p\alpha^2V_0^{\frac12}e^{-\frac{\alpha\phi}{2}}}{2\Gamma_0},
\end{equation}
and
\begin{equation}\label{etaH}
\eta_{_H}=\frac1r(\tilde{\eta}-\tilde{\epsilon})=\epsilon_{_H}.
\end{equation}
The condition of $\epsilon_{_H}=1$ is equivalent to $\ddot a=0$. From Eq. (\ref{epsilonH}), we can see that $\epsilon_{_H}$ decreases as the inflaton field rolls down the potential, which means that, after $\epsilon_{_H}$ is reduced to $1$, it will remain less than $1$ forever. Therefore, the condition of $\epsilon_{_H}=1$ implies the beginning of the inflation instead of the end of the inflation as usual, which is probably mistaken in \cite{TachyonWI}. Using Eq. (\ref{epsilonH}), we get the inflaton at the beginning of the inflation
\begin{equation}\label{phii}
\phi_i=\frac{1}{\alpha}\ln\left(\frac{3M_p^2\alpha^4V_0}{4\Gamma_0^2}\right),
\end{equation}
and the potential at the beginning of the inflation
\begin{equation}\label{potential}
V_i=\frac{4\Gamma_0^2}{3M_p^2\alpha^4}.
\end{equation}
Using Eq. (\ref{efold}), we get the total number of e-folds
\begin{eqnarray}
N &=&-\frac 1{M_p^2}\int_{\phi _i}^{\phi _e}\frac{V^2}{V,_\phi }rd\phi
\nonumber \\
&=&\frac{2\Gamma _0}{\alpha ^2}(3M_p^2V_0)^{-\frac 12}\left( e^{\alpha \phi
_e/2}-e^{\alpha \phi _i/2}\right) ,  \label{efold1}
\end{eqnarray}
Using Eq. (\ref{potential}), we rewrite the equation above in terms of $V_e$ and $V_i$ as
\begin{equation}\label{efold2}
N=\left(\frac{V_i}{V_e}\right)^{\frac12}-1.
\end{equation}
The number of the e-folds is no longer a small number even less than one as in \cite{TachyonWI,XiaoZhu} since $V_i\gg V_e$.

We should point out a serious problem with this. The parameter $\epsilon_{_H}$ is a decreasing function of $\phi$, and the inflaton $\phi$ gets larger as it rolls down its potential. Thus $\epsilon_{_H}$ is decreasing with time, and it will always be less than one after it passes one, which means the inflation will continue forever. This counters the evolutionary history of the Universe. Therefore, we reach the conclusion that the HSR parameter $\epsilon_{_H}$ must be an increasing function of time during the inflationary phase, thus $\epsilon_{_H}$ can increase to $1$ to end the inflation and the Universe can turn into the radiation dominated Big-Bang phase. A crucial requirement for an inflationary model is having an increasing HSR parameter $\epsilon_{_H}$. (There may be additional requirements, which deserve more research.) From the discussion we can see that $\Gamma=\Gamma_0$ is not a suitable model for the tachyon warm inflation with an exponential potential (which may be the model for dark energy). The failure of this model can be cured by assuming $\Gamma$ as a function of $\phi$.

The discussions above are restricted to the tachyon warm inflationary models. Now we try to extend the problem of the end of inflationary models to a broader scope. In a cold inflation scenario, there are a variety of models, including new, chaotic, power-law, hybrid, natural, brane, k. ghost, tachyon, etc. Generally speaking, most of them fall into one of four kinds: large-field models, small-field models, hybrid inflation and double inflation \cite{inflationdynamics}. Now we focus on the single-field inflationary models. Models such as the chaotic inflation (large-field models with a potential $V=\frac12 m^{2}\phi^2$) and the natural inflation (small-field models with a potential $V=m^4[1+\cos(\phi/f)]$) all have a potential minimum, and as a result the inflaton can oscillate about the minimum to end the inflation and heat up the Universe. However, some single-field models such as the quintessential
inflation and the tachyon inflation, do not have a potential minimum, and they need new efficient mechanism such as the instant preheating \cite{preheating} and the curvaton reheating \cite{curvatonreheating} for ending the inflation. The HSR parameter $\epsilon_{_H}$ may not be an increasing function in cold inflation, if we introduce some reheating mechanisms. The HSR parameter $\epsilon_{_H}$ for the cold tachyon inflation is
\begin{equation}\label{coldtachyon}
\epsilon_{_H}=\frac{M_p^2}{2}\frac{\alpha^2}{V_0}e^{\alpha\phi/2},
\end{equation}
which is an increasing function of time and consistent with the criterion obtained above. But the cold tachyon inflationary picture still suffers from some problems. The energy density of the tachyon field evolves in the form of $\rho_{\phi}\propto a^{-3\dot\phi^2}$, as the equation of state is $\omega=\dot\phi^2-1$. While $0<\dot\phi^2<1$, the radiation energy ($\rho_r\propto a^{-4}$) created at the end of the inflation would redshift faster than the energy density in the tachyon field \cite{reheatingproblem}, which is inconsistent with the radiation-dominated Big Bang Universe. Therefore, using a tachyon field as the inflaton field, the warm inflationary scenario with the decay of inflaton to thermal radiation is more applicable. But the condition of $\epsilon_{_H}$ being an increasing function should be satisfied since there is not a reheating period in warm inflation.

\subsection{\label{sec:level42}$\mathbf{\Gamma}$ as a function of $\mathbf{\phi}$}
As in \cite{TachyonWI,SetareKamali}, here we take the dissipative term of the form $\Gamma=f(\phi)=c^2V(\phi)=c^2V_0e^{-\alpha\phi}$, where $c^2>0$. In this case, the PSR parameters $\tilde{\epsilon}$, $\tilde{\eta}$ are the same as in the $\Gamma=\Gamma_0$ case, and
\begin{equation}\label{beta}
\tilde{\beta}=\frac{\alpha^2M_p^2}{V_0}e^{\alpha\phi}.
\end{equation}
The dissipative rate $r$ is
\begin{equation}\label{rater1}
r=\frac{c^2M_pe^{\alpha\phi/2}}{\sqrt{3V_0}}.
\end{equation}
In this case the HSR parameters $\epsilon_{_H}$ and $\eta_{_H}$ are given by
\begin{equation}\label{epsilonH1}
\epsilon_{_H}=\frac{\sqrt{3}\alpha^2M_pe^{\alpha\phi/2}}{2c^2V_0^{1/2}},~~~\eta_{_H}=\epsilon_{_H}.
\end{equation}
We find from the equation above that $\epsilon_{_H}$ is an increasing function and satisfies the requirement for an inflationary model. When $\epsilon_{_H}=1$, we get the inflaton at the end of inflation
\begin{equation}\label{phie}
\phi_e=\frac{2}{\alpha}\ln\left(\frac{2c^2V_0^{1/2}}{\sqrt{3}\alpha^2M_p}\right).
\end{equation}
Using Eq. (\ref{phie}), we obtain the potential at the end of inflation
\begin{equation}\label{Vend}
V_e=\frac{3\alpha^4M_p^2}{4c^4}.
\end{equation}
Thus the total number of the e-folds is given by
\begin{equation}\label{efold3}
N=\frac{2c^2V_0^{1/2}}{\sqrt{3}M_p\alpha^2}\left(e^{-\alpha\phi_i/2}-e^{-\alpha\phi_e/2}\right).
\end{equation}
Using Eq. (\ref{Vend}), we rewrite the equation above as
\begin{equation}\label{efold4}
N=\left(\frac{V_i}{V_e}\right)^{\frac12}-1.
\end{equation}
As long as $V_i>10^4V_e$, we can get enough number of e-folds. Using the expression of $V$, and set $N=60$ we have $\triangle\phi=\phi_e-\phi_i\approx 8/\alpha$. As what is estimated in \cite{TachyonWI}, we set $c^2=10^7M_p$ and $T\simeq T_r=10^{16}Gev$, and then we can estimate the parameters to be $\alpha\approx10^{-6}M_p$ and $V_{\ast}\approx10^{-10}M_p^4$ ($V_{\ast}$ is the potential when Hubble horizon crossing) using Eqs. (\ref{scalarperturbation1}) and (\ref{tensorperturbation}). Then we can get the field variation as $\triangle\phi\approx10^6M_P^{-1}$, and the field variation of the uniform normalized field as $\triangle\psi=\int\sqrt{V}d\phi\approx\sqrt{V_{\ast}}\triangle\phi\approx10 M_p$. The amplitude of field variation of $\psi$ can be near the order of Planck mass, which is much smaller than that of large-field models.

Now we rewrite Eq. (\ref{epsilonH1}) as
\begin{equation}\label{epsilonH2}
\epsilon_{_H}=\frac{3\alpha^2M_p^2}{2}\frac{H}{\Gamma}.
\end{equation}
We know that $H\propto \sqrt{V}\propto e^{-\alpha\phi/2}$. If we assume that $\Gamma$ has the form of $\Gamma\propto e^{-\beta \phi}$, then the condition $\beta>\alpha/2$ should be met in order not to violate the condition for a workable inflationary model. Detailed analysis of different microscopic models of interaction between canonical inflaton field and other fields suggested the dissipative coefficient may have a general form of $\tilde{\Gamma}\propto T^m\phi^n$ \cite{Lisa2004,Campo2010,ZhangZhu}, but this kind of pure power-law form for $\Gamma$ is not suitable for the case of tachyon warm inflationary scenario, for it will also give a decreasing $\epsilon_{_H}$. Other forms of $\Gamma\propto\tilde{\Gamma}(\phi,T)V(\phi)$ deserve more research. Based on the discussions above, we can check that, with an exponential potential, $\Gamma=c^2V(\phi)$ is a good choice while $\Gamma=\Gamma_0$ is not.

Now we analyze the power spectrum of a scalar perturbation and a tensor perturbation.

The scalar perturbation is given by \cite{TachyonWI}
\begin{equation}\label{scalarperturbation}
P_R=\frac{\sqrt{3}}{12\pi^2}\frac{\exp[-2\mathfrak{S}(\phi)]}{r^{1/2}\epsilon_{_H}}\frac{T_r}{H},
\end{equation}
where $\mathfrak{s}(\phi)$ is given by
\begin{eqnarray}
\mathfrak{S}(\phi)=-\int\left[\frac{1}{3Hr}\left(\frac{\Gamma}{V}\right),_{\phi}+\frac89\frac{V,_{\phi}}{V}\left(1-
\frac{(\ln\Gamma),_{\phi}(\ln V),_{\phi}}{36rH^2}\right)\right].\nonumber \\
\label{sphi}
\end{eqnarray}
Spectral index $n_s$ is calculated in \cite{TachyonWI}
\begin{equation}\label{indexn}
n_s\approx1-\left[\frac{3\eta_{_H}}{2}+\epsilon_{_H}\left(\frac{2V}{V,_{\phi}}\left[2\mathfrak{S},_{\phi}-
\frac{r,_{\phi}}{4r}\right]-\frac52\right)\right].
\end{equation}
The variables $\Gamma$, $T$ and $r$ all appear in the expressions of scalar power spectrum and spectral index, makeing these expressions complicated.
Using the equation above, we can obtain the spectral index in the form of
\begin{equation}\label{indexn1}
n_s-1\approx\frac74\frac{\tilde{\beta}}{r}-\frac54\frac{\tilde{\epsilon}}{r}-\frac32\frac{\tilde{\eta}}{r}.
\end{equation}
With the slow-roll conditions holding, we can get a nearly scale-invariant power spectrum.
The concrete form of the scalar perturbation in our case can be written out as
\begin{equation}\label{scalarperturbation1}
P_R(k_0)=\frac1{2\pi^2}\frac{T_rcV^{1/4}}{3^{1/4}M_p^{1/2}\alpha^2}exp\left(-\frac94\alpha\phi+\frac{\sqrt{3}\alpha^2M_p}
{8c^2V^{1/2}}\right),
\end{equation}
where Eq. (\ref{scalarperturbation}) has been used.

The tensor perturbation and the spectral index $n_g$ are given by \cite{TachyonWI}
\begin{equation}\label{tensorperturbation}
P_T=\frac{2}{M_p^2}\left(\frac{H}{2\pi}\right)^2\coth\left[\frac{k}{2T}\right]\simeq\frac{V}
{6M_p^2}\coth\left[\frac{k}{2T}\right],
\end{equation}
where the temperature $T$ in an extra factor $\coth\left[\frac{k}{2T}\right]$ denotes the temperature of a thermal background of gravitational wave \cite{Bhattacharya2006},
\begin{equation}\label{ng}
n_g=-2\frac{\tilde{\epsilon}}{r}.
\end{equation}
Using Eqs. (\ref{scalarperturbation1}) and (\ref{tensorperturbation}), we have the tensor-scalar ratio as
\begin{eqnarray}
R(k_0) &=&\left( \frac{P_T}{P_R}\right) _{k=k_0}=\frac{V^{3/4}\alpha ^2}{%
3^{3/4}M_p^{7/2}T_rc}  \nonumber \\
&&\times exp\left( -\frac 94\alpha \phi +\frac{\sqrt{3}\alpha ^2M_p}{%
8c^2V^{1/2}}\right) \coth \left[ \frac k{2T}\right] ,  \label{ratio}
\end{eqnarray}
where these quantities are valued when the scale $k_0=0.002Mpc^{-1}$ was leaving the horizon.

The scalar and tensor power spectrum and their index are quite different from both the tachyon cold inflation and the conventional warm inflation. It is not surprising, since the tachyon field has a very different Lagrangian density and equation of motion from non-tachyon scalar field and thermal fluctuations dominate the density perturbations.

\section{\label{sec:level5} Inverse power law potential in the strong dissipative regime}

In this section we consider another type of potential that also satisfies the two properties of the tachyon field potential: an inverse power-law potential
\begin{equation}\label{potential1}
V(\phi)=C\phi^{-m},
\end{equation}
where $C$ and $m$ are positive parameters.
This type of potential also does not have a minimum ($V\rightarrow0$ while $\phi\rightarrow\infty$ ), and can result in a tachyon warm-intermediate inflation \cite{SetareKamali}. The inverse power-law potential was introduced in \cite{Peebles} and was studied as a model for dark energy. Here it is used as a model for tachyon warm inflation. The case of $m=2$ has been analyzed in standard tachyon model \cite{HuangZhu} and the case of $m=4$ has been studied using the Noether gauge symmetry in $f(R)$ tachyon model \cite{Jamil2011}. In this section, again we focus on the strong dissipative regime ($r\gg1$).

\subsection{\label{sec:level51}$\mathbf{\Gamma=\Gamma_0}$ case.}
With $\Gamma=\Gamma_0=constant$, and using the inverse power-law potential given by Eq. (\ref{potential1}), we obtain the PSR parameters as
\begin{equation}\label{PSRparameter}
\tilde{\epsilon}=\frac{M_p^2m^2}{2C}\phi^{m-2},~~~\tilde{\eta}=\frac{M_p^2m(m+1)}{C}\phi^{m-2},~~~\tilde{\beta}=0.
\end{equation}
The other slow-roll parameters are $b=0$ and $c=0$, as in the exponential potential case. The Hubble parameter and the dissipative rate $r$ are given by
\begin{equation}\label{Hubble1}
H(\phi)=\sqrt{\frac C3}\frac{\phi^{-m/2}}{M_p},~~~
\end{equation}
and
\begin{equation}\label{rater2}
r=\frac{\Gamma_0M_p\phi^{3m/2}}{\sqrt{3}C^{3/2}}.
\end{equation}
The HSR parameter $\epsilon_{_H}$ and $\eta_{_H}$ are given by
\begin{equation}\label{epsilonH3}
\epsilon_{_H}=\frac{\sqrt{3C}M_pm^2}{2\Gamma_0\phi^{\frac12 m+2}},~~~\eta_{_H}=\frac{\sqrt{3C}M_pm(m+2)}{\Gamma_0\phi^{\frac12m+2}}.
\end{equation}
From Eq. (\ref{epsilonH3}), we can see that, in the $\Gamma=\Gamma_0$ case, $\epsilon_{_H}$ is also a decreasing function in violation of the condition for a workable inflationary model, thus making $\Gamma=\Gamma_0$ not suitable for the inverse power-law potential (the tachyon warm-intermediate inflation case proposed in \cite{SetareKamali}). The potential of the intermediate inflation does not have a minimum either, and the HSR parameter for the cold non-tachyon intermediate inflation is $\epsilon_{_H}=M_p^2m^2/2\phi^2$, which is a decreasing function of time. Therefore, the cold non-tachyon intermediate inflationary models also need some mechanism for ending the inflation. In the tachyon cold intermediate inflation \cite{tachyonintermediate}, the HSR parameter $\epsilon_{_H}=M_p^2m^2\phi^{m-2}/2C$, and, if $m>2$, the inflation can end naturally when $\epsilon_{_H}=1$. But the cold tachyon suffers from the problem of redshifting slower than the radiation after inflation, as we mentioned in the previous section.

Now we rewrite Eq. (\ref{epsilonH3}) in the form of
\begin{equation}\label{epsilonH4}
\epsilon_{_H}=\frac{\sqrt{3C}M_pm^2\phi^{-2-\frac m2}}{2\Gamma}.
\end{equation}
If we assume $\Gamma\propto\phi^{-\beta}$, then the condition of $\beta>2+m/2$ should be met in order to have an increasing $\epsilon_{_H}$. Next we shall calculate the case of $\Gamma$ being a function of $\phi$ in order to give a self-consistent tachyon warm inflationary model with an inverse power law potential.

\subsection{\label{sec:level52}$\mathbf{\Gamma}$ as a function of $\phi$.}

We take the dissipative coefficient to be of the form of $\Gamma=g(\phi)=f^2V(\phi)=f^2C\phi^{-m}$ \cite{SetareKamali}, where $f$ is a constant and $f^2>0$. The inverse power-law form of dissipative coefficient ($\tilde{\Gamma}\propto\phi^{-n}$) was proposed in \cite{Zhangyi} based on some different microphysical basis of canonical scalar field. We shall check whether it will apply to the tachyonic case. The PSR parameters, $\tilde{\epsilon}$ and $\tilde{\eta}$, are given by Eq. (\ref{PSRparameter}), and $\tilde{\beta}$ is
\begin{equation}\label{beta1}
\tilde{\beta}=\frac{M_p^2m^2\phi^{m-2}}{C};
\end{equation}
the Hubble parameter is given by (\ref{Hubble1}), and the dissipative rate $r$ is
\begin{equation}\label{rater3}
r=\frac{M_pf^2}{\sqrt{3C}}\phi^{m/2}.
\end{equation}
The HSR parameters $\epsilon_{_H}$ and $\eta_{_H}$ in this case are given by
\begin{equation}\label{epsilonH5}
\epsilon_{_H}=\frac{\sqrt{3}M_pm^2}{2f^2\sqrt{C}}\phi^{\frac m2-2},~~~\eta_{_H}=\frac{\sqrt{3}M_pm(m+2)}{2f^2\sqrt{C}}\phi^{\frac m2-2}.
\end{equation}
If $m>4$, $\epsilon_{_H}$ can be an increasing function. Under that assumption, we get the inflaton at the end of inflation using Eq. (\ref{epsilonH5}),
\begin{equation}\label{phie1}
\phi_e=\left(\frac{2\sqrt{C}f^2}{\sqrt{3}M_pm^2}\right)^{\frac{2}{m-4}}.
\end{equation}
Using the equation above we can obtain the total number of e-folds,
\begin{equation}\label{efold5}
N=\frac{m}{m-4}\left[\left(\frac{V_i}{V_e}\right)^{\frac{m-4}{2m}}-1\right].
\end{equation}
For a concrete example, when $N=60$ and $m=6$, we have $V_i\simeq 10^7 V_e$, and thus $\phi_e\simeq10^{7/6}\phi_i$.

There is also a problem in this argument. As stated in \cite{SetareKamali}, the tachyon field with an inverse power law potential can result in a warm-intermediate inflation, where the scale factor has the form of
\begin{equation}\label{scalefactor}
    a(t)=a_0exp(At^k),  ~~~0<k<1.
\end{equation}
In our case, $k=(4-m)/4$. To have for $k>0$ (the Universe is in an inflationary phase), we need $m<4$, which violates the condition of $\epsilon_{_H}$ being an increasing function. Hence, the model of $\Gamma=f^2V(\phi)$ with an inverse power-law potential proposed in \cite{SetareKamali} is not a workable one either.

If we assume the dissipative coefficient as having the form of $\Gamma\propto\phi^{-n}$, then
\begin{equation}\label{epsilonHend}
\epsilon_{_H}=\frac{\sqrt{3C}M_pm}{2}\phi^{n-m/2-2},
\end{equation}
and $k=(m/2-n+2)/(m-n+2)$ in Eq. (\ref{scalefactor}). $\epsilon_{_H}$ being an increasing function and $k>0$ implies that $n>m+2$. The exponential form of dissipative coefficient seems to be unreasonable here, and shall be left out of our consideration.

So far we check and obtain the consistency conditions for a workable tachyon warm-intermediate inflationary models.

\section{\label{sec:level6}Discussions and Conclusions}

In this paper we investigate the consistency of the tachyon warm inflationary models. The paper begins with a short review of the tachyon field and a brief introduction to the tachyon warm inflationary Universe. Since inflation of Universe is often associated with a slow-roll solution, we introduce parameters for the slow-roll inflation, which are divided into two kinds: the Hubble slow-roll parameters ($\epsilon_{_H}$ and $\eta_{_H}$) in the Hamilton-Jacobi form, and the potential slow-roll parameters ($\tilde{\epsilon}$, $\tilde{\eta}$ and $\tilde{\beta}$) which have obvious relationship with the inflaton potential. The HSR parameters are often used in the study of the tachyon warm inflationary models. In this paper we redefine some new PSR parameters that are different from the non-tachyon scalar field, since the tachyon field has a non-canonical kinetic term in their Lagrangian density. Two other slow-roll parameters are introduced to describe the temperature dependence of the potential and the dissipative coefficient. The validity of the slow-roll approximation requires that the slow-roll solution act as an attractor for the dynamical system. We perform a linear-stability analysis to find the conditions for the validity of the slow-roll solution. The stability analysis can be written in a concise form using the new PSR parameters. Our analysis yields the slow-roll conditions: $\tilde{\epsilon} \ll 1+r$, $|\tilde{\eta}|\ll 1+r$, $|\tilde{\beta} |\ll 1+r$ and $|b|\ll  r/(1+r)$, $|c|<4$. The first three slow-roll conditions are different from that in the canonical scalar field \cite{Ian2008}, but have the same form after using the newly defined PSR parameters. During the slow roll inflationary regime, tachyon field acts as a canonical field but evolves quite differently at late times. The conditions for $b$ and $c$ are the same as those in the canonical scalar field, i.e., the thermal correction to the effective potential should be negligible and the temperature dependence of the dissipative coefficient should be inside the range of $\Gamma\propto (T^{-4}, T^4)$.

With the slow-roll conditions obtained, we study two cases of potential for the tachyon field: an exponential potential ($V(\phi)=V_0\exp(-\alpha\phi)$) and an inverse power-law potential ($V(\phi)=C\phi^{-m}$). Both potentials have the two properties of the tachyon field potential. In both studies, we consider a constant dissipative coefficient and it being a function of the inflaton. Correcting what we believe are mistakes made in previous papers, we obtain the number of e-folds in the $\Gamma=\Gamma_0$ case with an exponential potential. Through the analysis for the examples, we propose a crucial condition for the workable inflationary models: the HSR parameter $\epsilon_{_H}$ should be increasing as the inflaton rolls down its potential. Otherwise, the inflation of Universe will last forever and destroy the thermal history of our Universe, unless some new mechanism ends the inflation and heats up the Universe in a cold inflation. We also extend our discussion to a broader range of inflationary models. In a warm inflation, the criterion must be satisfied since there is no reheating period and the connection to radiation-dominated phase is smooth. For both potentials in the tachyon warm inflationary case, we find $\Gamma=\Gamma_0=const.$ to not be a suitable choice or a reasonable assumption, as it gives a decreasing $\epsilon_{_H}$ for both potentials, despite the benefit of simplifying calculations. We also give out the conditions that the dissipative coefficient should satisfy for both potentials. Based on the conditions obtained, we analyze the density fluctuations with an exponential potential, and we get a nearly scale-invariant power spectrum.

There could be other consistent conditions for the tachyon warm inflationary Universe models, which deserve more research.

\acknowledgments This work was supported by the National Natural Science Foundation of China (Grant No. 11175019 and No. 11235003).

\end{document}